\documentclass[structabstract]{aa}  

\usepackage{graphicx}
\usepackage{times,epsfig}
\usepackage[intlimits,sumlimits]{amsmath}  
\usepackage[dvips]{color}  
\usepackage{natbib}
\bibpunct{(}{)}{;}{a}{}{,}
\usepackage{hyperref}
\usepackage{longtable}
\usepackage{lscape}
\usepackage{verbatim}
\usepackage{txfonts}
\usepackage{amssymb}
\usepackage{mathrsfs}  
\usepackage{epstopdf}
\usepackage[english]{babel}
\usepackage{url}

\newcommand{\beqa}{\begin{eqnarray}} 
\newcommand{\eeqa}{\end{eqnarray}}

\newcommand{\bsub}{\begin{subequations}}
\newcommand{\esub}{\end{subequations}}
\newcommand{\beal}{\begin{align}}
\newcommand{\ealn}{\end{align}}

\begin{document}

\title{The properties of SN Ib/c locations\thanks{Based on observations collected at the European Southern Observatory during ESO Program 081.D-0757(A). }}
\titlerunning{The properties of SN Ib/c locations}

\author{
G.~Leloudas\inst{1} 
\and A.~Gallazzi\inst{1}
\and J.~Sollerman\inst{2} 
\and M.~D.~Stritzinger\inst{2}
\and J.~P.~U.~Fynbo\inst{1}
\and J.~Hjorth\inst{1}
\and D.~Malesani\inst{1}
\and M.~J.~Micha\l owski\inst{3}
\and B.~Milvang-Jensen\inst{1}
\and M.~Smith\inst{4}
}

\institute{Dark Cosmology Centre, Niels Bohr Institute, University of Copenhagen, Juliane Maries Vej 30, 
2100 Copenhagen \O, Denmark
\and The Oskar Klein Centre, Department of Astronomy, Stockholm University, AlbaNova, 10691 Stockholm, Sweden
\and Scottish Universities Physics Alliance, Institute for Astronomy, University of Edinburgh, Royal Observatory, Edinburgh, EH9 3HJ,~UK
\and Astrophysics, Cosmology and Gravity Centre, Department of Mathematics and Applied Mathematics, University of Cape Town, Cape Town 7700, South Africa
}

\offprints{\email{giorgos@dark-cosmology.dk}}
\date{Received 10 February 2011 / Accepted 7 March 2011}

\abstract{} 
{
We seek to gain a deeper understanding of stripped-envelope, core-collapse supernovae through studying their environments.
}
{
We obtained low-resolution optical spectroscopy with the New Technology Telescope ($+$ EFOSC2)  at the locations of 20 type~Ib/c supernovae.
We measured the flux of emission lines in the stellar-continuum-subtracted spectra from which local metallicities 
are computed. 
For the supernova regions, we estimate both the mean stellar age, by interpreting the stellar absorption with population
synthesis models, and the age of the youngest stellar populations using  the H$\alpha$ equivalent
width as an age indicator. 
These estimates are compared with the lifetimes of single massive stars.
} 
{Based on our sample, we detect a tentative indication that type~Ic supernovae might explode in environments that are more metal-rich than those of  type~Ib supernovae (average difference of 0.08 dex), but this is not a statistically significant result. 
The lower limits placed on the ages of the supernova  birthplaces  are generally young, although there are several cases where these appear older than what is expected for the evolution of single stars that are more massive than 25--30 $M_{\sun}$. 
This is only true, however, when assuming that the supernova progenitors were born during an instantaneous (not continuous) episode of star formation. 
}
{
These results do not conclusively favor any of the two evolutionary paths (single or binary) leading to stripped supernovae.
We do note a fraction of events for which binary evolution is more likely, due to their associated age limits;
however, the  supernova environments contain areas of recent ($<$15 Myr) star formation, and  the environmental metallicities  at least 
do not contradict the single evolutionary scenario, suggesting  that this channel is also broadly consistent with the observations.
}


\maketitle

\section{Introduction}

The progenitors of stripped-envelope core-collapse (CC) supernovae (SNe) still evade direct detection \citep{smarttARAA}, although reasoning dictates that  they are Wolf-Rayet (WR) stars \citep{WRreview}.
This class of explosions encompasses the hydrogen-deficient supernovae of types~Ib and Ic (collectively SNe~Ib/c), but also SNe~IIb that retain part of their hydrogen envelope. 
In \cite{LeloudasWRloc} we showed that the locations of SNe~Ib/c
within their host galaxies  are indeed consistent with those of WR stars.
Several open questions exist in relation to these stellar explosions, the most fundamental of which probably is 
whether they result from the evolution of single or binary stars.

To appear as a SN~Ib/c, the progenitor star has to undergo significant 
stripping of its outer hydrogen envelope. 
This could be the result of mass loss due to stellar winds in single massive stars.
Alternatively, the outer envelope could be shed owing to binary evolution \citep[e.g.][]{Podsial92}.
The idea that at least a significant fraction of stripped-envelope SNe come from a binary channel 
 presently enjoys significant support 
\citep[e.g.][]{2007PASP..119.1211F,smarttARAA,2011MNRAS.412.1522S},
based largely on arguments related to the stellar initial mass function (IMF) and the observed ratio of type~Ib/c to type~II SNe.

Valuable insight into the
nature of these explosions can be gained through observations of their environments
because the environmental properties can be used for a number of tests.
In the single massive star
evolutionary scenario, normal SNe~Ic are expected to arise in higher
metallicity environments than SNe~Ib. This is because wind-driven mass loss for single massive stars depends strongly on metallicity \citep{vink},
and SNe~Ic need to lose more mass than SNe~Ib.
If the mass loss is instead caused by binary evolution, it is probably independent of metallicity \citep{2007PASP..119.1211F}.  
Therefore, determining the metallicity at the locations
of type Ib/c SNe has the potential of probing their origins.
Another potential probe of the explosion progenitor is to
determine the age of the local
stellar population. 
This is because 
single massive stars should only be found in young, actively star-forming
regions. 
For such environmental studies, large
samples are necessary for a statistically reliable approach.

Although indirect \citep{PB03} and direct \citep{Prieto2008} 
global host galaxy metallicity measurements exist, studies
focusing on the local environments have been based  until recently on
proxies of metallicity \citep{AndJames2009,BP09}.  
Two contemporary studies \citep{andersonCCmetal,modjazIbcMetal}
 report on SN host galaxy 
metallicity measurements that are both local (at the site of the SN) and direct 
(using local emission-line metallicity estimates).
\cite{andersonCCmetal} studied 74 host H\,II regions of CC SNe,  of which 27
were type~Ib/c SNe, while the \cite{modjazIbcMetal} study focuses on 35 SNe~Ib/c. 
The two studies  disagree somewhat regarding the existence of 
differences in the environmental metallicities between type Ib and type Ic SNe.
 \cite{andersonCCmetal} find equal metallicities between the two,
while \cite{modjazIbcMetal} report on a difference of 0.20 dex, which they argue is statistically significant.

In this paper we report on similar observations of 20 SNe Ib/c sites (15 of which are new),
thereby increasing the existing sample. Our targets, observations, and
data reduction are presented in the next section, where we also discuss
the critical issue of  spectroscopic classifications.
In Sect.~\ref{sec:results} the methods for subtracting the stellar continuum from the spectra, measuring the  local metallicities at the SN sites
and estimating the age of the local stellar populations are described.
Section~\ref{sec:disc} contains a discussion of our results and Sect.~\ref{sec:conc} concludes our study.

\section{The data and data reductions}
\label{sec:data}

We  targeted host galaxies of normal stripped-envelope CC SNe 
that have been
closely monitored by either the Carnegie Supernova Project (CSP) 
or  the SDSS-II Supernova Survey. 
For the present study the sample selection is only 
relevant for understanding the potential biases in our data.

The CSP  \citep{2006PASP..118....2H} obtained unprecedented photometry and spectra 
for $\sim$250 SNe of all types, 35 of which were stripped CC SNe. 
These SNe were discovered by different search programs that typically monitor 
bright ($m_B \sim 11-16$) nearby galaxies. This, of course, introduces an 
unavoidable selection bias.

The SDSS-II Supernova Survey 
\citep{2008AJ....135..338F} scanned an equatorial strip of the sky  
searching  for SNe~Ia to be used for cosmology
\citep[see][]{2009ApJS..185...32K}. 
Other types of SNe were also discovered, 
including $\sim$20 stripped-envelope CC SNe.
The SDSS SN host galaxies are typically fainter 
($m_g \sim 17-22$) and less spatially resolved than those followed by the CSP. 
Spectra for some of the brightest of these galaxies exist in the SDSS database 
\citep[e.g.][]{Prieto2008}, but these spectra are centered on the 
galaxy nucleus and not on the SN location itself.
The SDSS sample is a relatively unbiased sample from a host galaxy point of view.
We note, however, that this survey was targeting mainly SNe~Ia and that the selection of CC~SNe 
was not made in order to obtain a complete, unbiased sample.
Some CC~SNe were only observed spectroscopically at lower priority, 
depending on the availability of telescope time, while other events were initially (photometrically) misidentified as SNe~Ia.

The data presented in this paper were collected during three consecutive nights (21 -- 23 August 2008) 
with the ESO New 
Technology Telescope (NTT) equipped with EFOSC2.  
The galaxies observed were those  that matched the  RA and DEC window best,
so this should not introduce any more biases.  There is no overlap
with the sample of  \cite{andersonCCmetal} and only a small overlap (5
SN hosts) with the sample of \cite{modjazIbcMetal}.  
The \cite{modjazIbcMetal} sample also includes SNe that were
discovered in `non-targeted' searches (such as the SDSS-II SN Survey)
while the \cite{andersonCCmetal} hosts are mostly `targeted' galaxies.

A list of the observed galaxies is provided in
Table~\ref{tab:galaxies}.  This table contains basic information on the
host galaxy, the hosted SN, as well as details
concerning our observations.  We comment further on the critical issue of SN
typing in Sect.~\ref{subsec:clas}.  
The tabulated offsets are SN galactocentric distances (projected on the plane of  the sky), based on the redshift of the host galaxies.
The following cosmological parameters have been adopted throughout the paper:
H$_{\rm 0}=$73~km~s$^{-1}$~Mpc$^{-1}$, $\Omega_{\rm M}=0.27$, and $\Omega_{\rm \Lambda}=0.73$.
For the two galaxies that are not in the Hubble flow (i.e. NGC~1187
and NGC~4981), we  used the distances reported by
\cite{2009AJ....138..323T}.  For the solar metallicity, we adopted the following value:  $\log\rm{(O/H)}+12 = 8.69$ \citep{2009ARA&A..47..481A}.

\begin{table*}
\tiny
\setlength{\tabcolsep}{1.5 pt}
\caption[]{SN host galaxy sample: properties and observing log.}
\label{tab:galaxies}
\centering
\begin{tabular}{llccccccclll}
\hline\hline
Galaxy		           &  SN       &   RA (J2000)\tablefootmark{a}	& DEC (J2000)\tablefootmark{a}	  &   Offset\tablefootmark{b} & $M_B$\tablefootmark{c} &  Redshift\tablefootmark{d}	& type\tablefootmark{e}  &  Sample\tablefootmark{f}     &  Resolution  &  Exposure time			& Airmass       \\	
		           &         &   ($\degr$  $\arcmin$  $\arcsec$) & ($\degr$  $\arcmin$  $\arcsec$)	  &   (kpc) &   &  	&    &      & (\AA$/1\arcsec$)   &  (sec)		   &       \\	
\hline
2MASXJ21024677$-$0405233   &  2007hn   & 21 02 46.85	& $-$04 05 25.2  &  	1.06   & $-$20.35 & 0.0273  	 &  Ic	  & CSP    (N) &  13.6	     &  4$\times$1800			& 1.14\\ 
ESO 153$-$G17	      	   &  2004ew   & 02 05 06.17	& $-$55 06 31.6  &  	5.40   & $-$20.98 & 0.0218	 &  Ib    & CSP    (T) &  17.2       &  1800				& 1.13        \\ 
ESO 552$-$G40	      	   &  2004ff   & 04 58 46.19	& $-$21 34 12.0  &  	5.47   & $-$21.05 & 0.0226	 & Ib/IIb & CSP    (T) &  17.2       &  1800				& 1.11        \\ 
IC 4837A	      	   &  2005aw   & 19 15 17.44	& $-$54 08 24.9  &  	6.11   & $-$21.60 & 0.0094	 &  Ic    & CSP    (T) &  17.2, 8.2  &  2$\times$1200, 2$\times$1800	& 1.13, 1.11   \\ 
J000109.19$+$010409.5      &  2007nc   & 00 01 09.30	& $+$01 04 06.5  &  	8.76   & $-$19.83 & 0.0860	 &  Ib    & SDSS   (N) &  17.2       &  2$\times$1800		       & 1.44	   \\
J001039.34$-$000310.4      &  2007sj   & 00 10 39.63	& $-$00 03 10.2  &  	3.54   & $-$20.88 & 0.0390	 &  Ic    & SDSS   (N) &  17.2       &  2$\times$1800		       & 1.25	   \\
J002741.89$+$011356.6      &  2007qx   & 00 27 41.78	& $+$01 13 59.6  &  	5.63   & $-$20.02 & 0.0800	 &  Ic    & SDSS   (N) &  17.2       &  2$\times$1800		       & 1.18	   \\
J012314.96$-$001948.8      &  2006jo   & 01 23 14.71	& $-$00 19 46.7  &  	6.79   & $-$20.59 & 0.0770	 &  Ib    & SDSS   (N) &  17.2       &  2$\times$1800		       & 1.16	   \\ 
J023239.17$+$003700.1      &  2006fo   & 02 32 38.89	& $+$00 37 03.0	 &  	2.39   & $-$19.76 & 0.0201	 &  Ib    & both   (N) &  17.2       &  2$\times$1800			& 1.16        \\ 
J205121.43$+$002357.8      &  2007jy   & 20 51 21.43	& $+$00 23 57.8  &   \mbox{\ldots}\tablefootmark{g}    & $-$19.61 & 0.1831	 &  Ib    & SDSS   (N) &  16.8, 7.0  &  2100, 2$\times$1200	       & 1.45, 1.21  \\ 
J205519.76$+$003234.4      &  2005hl   & 20 55 19.79	& $+$00 32 34.8  &  	5.37   & $-$19.76 & 0.0230	 &  Ib    & SDSS   (N) &   8.2       &  3$\times$1500			& 1.18  	      \\ 
J213900.63$-$010138.6      &  2005hm   & 21 39 0.64	& $-$01 01 38.6	 &   \mbox{\ldots}\tablefootmark{g}    & $-$14.89 & 0.0347	 &  Ib    & SDSS   (N) &  17.2       &  2100			       & 1.30	   \\ 
J223529.00$+$002856.1      &  2007qw   & 22 35 29.01	& $+$00 28 56.2  &   \mbox{\ldots}\tablefootmark{g}    & $-$19.06 & 0.1507	 &  Ia    & SDSS   (N) &  16.8, 17.2 &  1800, 1800		       & 1.30	   \\ 
KUG 2302$+$073	      	   &  2006ir   & 23 04 35.68	& $+$07 36 21.5  &  	1.79   & $-$16.95 & 0.0200	 &  Ic    & CSP    (N) &  17.2       &  695				& 1.41        \\ 
MGC$+$03$-$43$-$5	   &  2005bj   & 16 49 44.74	& $+$17 51 48.7  &  	5.83   & $-$20.01 & 0.0222	 &  IIb   & CSP    (T) &  17.2       &  2$\times$1800			& 1.50        \\     
NGC 1187    	      	   &  2007Y    & 03 02 35.92	& $-$22 53 50.1  &  	9.83   & $-$20.17 & 0.0046	 &  Ib    & CSP    (T) &  17.2       &  2$\times$1800			& 1.07        \\ 
NGC 214     	      	   &  2006ep   & 00 41 24.88	& $+$25 29 46.7  &     13.44   & $-$21.62 & 0.0151	 &  Ib    & CSP    (T) &  17.2       &  2$\times$1800			& 1.74        \\ 
NGC 4981    	      	   &  2007C    & 13 08 48.80	& $-$06 46 45.0  &  	2.79   & $-$20.25 & 0.0056	 &  Ib/c  & CSP    (T) &  17.2       &  2$\times$1800			& 2.06        \\ 
NGC 7364    	      	   &  2006lc   & 22 44 24.48	& $-$00 09 53.5	 &  	3.14   & $-$21.21 & 0.0162	 &  Ib/c  & both  (TN) &  17.2       &  1800				& 1.25        \\ 
NGC 7803    	      	   &  2007kj   & 00 01 19.58	& $+$13 06 30.6  &  	4.15   & $-$20.88 & 0.0178	 &  Ib/c  & CSP    (T) &  17.2. 8.2  &  1200, 1800			& 1.42        \\ 
\hline\hline
\end{tabular} \\
\tablefoot{The order in the table is alphanumeric. Galaxies that were observed with more than one grism have double entries in the resolution, exposure time and airmass columns.
\tablefoottext{a}{RA and DEC of the SN.}
\tablefoottext{b}{Galactocentric distance of the SN region projected on the plane of the sky (at the distance of the host galaxy).}
\tablefoottext{c}{Galaxy absolute magnitude.}
\tablefoottext{d}{drawn from the NASA Extragalactic Database (NED) and SDSS, except for the hosts of SNe~2007hn, 2007nc, 2007jy, 2005hm and 2007qw which have redshifts determined from our spectroscopic observations.}
\tablefoottext{e}{See the detailed discussion in Sec.~\ref{subsec:clas}.}
\tablefoottext{f}{whether the SN belongs to the CSP or  the SDSS subsample with the discovery method noted in parentheses: T stands for targeted and N for non-targeted searches.}
\tablefoottext{g}{either point sources, or the SN location coincides with the galaxy nucleus.}
}
\end{table*}

Our observing strategy involved positioning the slit to contain both
the explosion site and the galaxy center.  This was done at the
expense of not observing at parallactic angle, but most objects were
observed at low airmasses (Table~\ref{tab:galaxies}) so any differential slit losses should be
minimal.  In most cases we used the EFOSC2 grism 11, but other grisms
were also used  to optimize
the resolution in the (restframe) wavelength region $3700-6800$~\AA.
A 1\arcsec \  wide slit
was used throughout the observing run.
It is noted that at the time of observations there was no significant contribution to the measured flux by  the underlying SNe.
SN~2007kj, the brightest of the SNe discovered in the fall of 2007 in our sample, is not expected to contribute with  more than 
4$\times$10$^{-19}$~erg~s$^{-1}$~cm$^{-2}$~\AA$^{-1}$ in the $V$-band  \citep[assuming a late-time evolution similar to the type~Ib SN~2007Y;][]{max07Y}.

All frames were bias-subtracted and flat-fielded with standard tasks
in IRAF\footnote{IRAF is distributed by the National
Optical Astronomy Observatory: \url{http://iraf.noao.edu/iraf/web/}.}.  
Cosmic rays were removed from the science spectra with 
the task {\tt LACosmic} \citep{lacosmic}.  The tasks {\tt
identify}, {\tt reidentify}, and {\tt fitcoords} were used on the HeAr
arcs  to create 2D dispersion maps and all 2D spectra were
wavelength calibrated with the  
task {\tt transform}.  We
extracted spectra using {\tt apall} both at the SN location
and at the galaxy nucleus.  For the SN location we tried to use the
smallest aperture possible (e.g. $7-10$ pixel wide columns) to minimize
contamination from neighboring regions.  
In some cases this still
corresponds to integrated light from areas on the order of 1 kpc$^2$,
in particular when the effect of seeing is included.  This
illustrates the limitations of this method, i.e. that the `local' SN
environment probed is in reality a large region, although this
method still probes the SN environment better than a nuclear
galaxy spectrum.  In four cases where no useful signal could be recovered at
the SN location, we resorted to extracting spectra of other regions,
which were as nearby as possible to the exact location and
which contained sufficient signal. 
The projected distances on the plane of the sky between the reported regions and the SN locations, at the distance of these galaxies
(that all appear face-on), are:
4.63 kpc for MGC$+$03$-$43$-$5, 4.11 kpc for NGC~1187, 7.19 kpc for NGC~214 and 2.42 kpc for NGC~7803.
In all cases we first traced the nuclear galaxy spectrum and then used the same trace function for all other apertures.

During our observing run, the
Moon was relatively bright and close to many of our targets.
The influence of the Moon was manifested as a halo in
the blue part of the CCD. This complicated pattern proved
difficult to correct for, causing problems in the background removal,
especially below 4000 \AA.  Combined with the reduced sensitivity of
the CCD in the blue, data below this wavelength therefore have
large associated uncertainties.

\subsection{SN classifications}
\label{subsec:clas}

Traditionally, SNe are classified by their optical
spectra around maximum light, and the distinction between SNe~Ib and SNe~Ic
is based on the relative strength of \ion{He}{i} and
\ion{O}{i} lines \citep{1997ARA&A..35..309F,2001AJ....121.1648M}.
The distinction between these subtypes can, however, be a subtle and
difficult matter.  

Following the temporal evolution of the spectra is usually a great
help, especially since the He lines in SNe~Ib 
typically become more prominent as
time passes, and it is not unusual that SNe are reclassified 
as multi-epoch spectra are obtained. 
In some cases, however,  the data coverage
and/or quality of the spectrum used to make the classification is 
poor so a subclassification is difficult, if not impossible. 
In fact, SNe~Ib/c form a continuum, depending on the amount of He that has been stripped off their progenitor.  
The existence of intermediate objects also consists a challenge for the subclassification scheme.

Great caution should be taken when adopting the reported SN types,
especially when the distinction between types~Ib and Ic 
 is important.  
The spectral classifications that are reported in
Table~\ref{tab:galaxies} are the result of careful inspection of the
spectra that were available to us, and in many cases they differ from what
is reported in the IAU or CBET circulars or in the continuously updated 
Asiago SN catalog
\citep{agiago_original}.  
The number of reclassifications we had to make 
is significant (8 out of 20 events). 
This shows that, if proper
care is not taken, even the statistical power of large  samples might be
affected.  The individual cases are briefly discussed below.

For the CSP subsample we note the following differences:  SNe~2006ir
and 2007hn were discovered and reported as SNe~Ib/c by the SN Factory
\citep{2006CBET..658....1P,2007CBET.1050....1B}. Based on multi-epoch
CSP spectra, we refine the classication to SNe~Ic.  
SN~2005bj is a SN~IIb and
not a SN~Ic as the strong absorption at $\sim$6350 \AA\ is clearly
due to H$\alpha$ (possibly blended with He $\lambda$6678) and not to 
Si as initially suspected \citep{2005CBET..138....1M}. 
SNID \citep{2007ApJ...666.1024B} also shows an excellent overall agreement
with other type~IIb objects.  Similarly, SN~2004ff is a SN~Ib (probably
even IIb) and not of type Ic, based on prominent He lines detected in a series
of 3 spectra. The first CSP spectrum is dated almost 2
weeks after the reported classification spectrum
\citep{2004IAUC.8428....3M}, illustrating 
the importance of temporal coverage.  For the purposes of this
paper, SNe~IIb will be grouped together with SNe~Ib.  For SNe~2006lc
and 2007kj we note  
a similarity with SN~1999ex \citep[see also][]{2006CBET..699....1B,2007CBET.1093....1O}, for which an
intermediate classification (Ib/c) is often quoted since its He lines are
weaker than in prototypical type Ib objects \citep{2002AJ....124..417H}. 
We have
attributed them such a transitional classification, although
we also discuss the consequences of including them in the type Ib sample.
SN~2007C is even less of a clear case and for
this reason we preferred a mixed Ib/c classification. 
The CSP
SNe~Ib/c spectra will be published in Stritzinger et al. (in prep.).

All spectra obtained at the NTT or the NOT for the SDSS-II SN Survey
have been carefully reduced and examined again in a comprehensive and
systematic way by \cite{2011A&A...526A..28O}, and are now publicly
available\footnote{\url{www.physto.se/~linda/spectra/nttnot.html}}.
These include 6 of the 9 SNe~Ib/c from our SDSS subsample.  In general, the
SDSS-II discovered SNe have spectra with lower S/N, suffer 
significant host galaxy contamination and typically span only $1-2$ epochs,
making classification a \textit{de facto}  more difficult task
compared to the nearby CSP SNe.  An exception is SN~2006fo, for which 
we have many spectra, including three from the CSP. The examination of
this event by \cite{2011A&A...526A..28O} led to the confident revision of its
type to Ib (from Ic), which is  now also the official SDSS-II
classification.  We  inspected the spectra of the other events and
conclude that the most probable types are indeed those that are
adopted by SDSS-II, and we  therefore retain these classifications.  
Admittedly, a detailed division between types~Ib and Ic  SN based on
these spectra might be overly ambitious. A more conservative approach is
followed by \cite{2011A&A...526A..28O} who prefer a general SN~Ib/c
classification for 2 events (SNe~2006jo and 2007qx), while for
SNe~2007sj and 2007jy, they conclude that even a type~Ib/c classification
is not unambiguous.  Finally, there is also the case of SN~2007qw,
whose spectral type has been revised from type~Ib to type~Ia by the SDSS-II
collaboration (M.~Sako; priv. comm.). We have
therefore only included it in our tables and not in the analysis.\footnote{After the present paper was accepted, new communication revealed that the SDSS-II classification for this SN might be revised again. \cite{modjazIbcMetal} include it in their broad-lined SN~Ic sample. In any case, it is probably better to exclude it due to the related uncertainties.}

\section{Methods and results}
\label{sec:results}

\begin{figure}
\begin{center}
\includegraphics[width=\columnwidth]{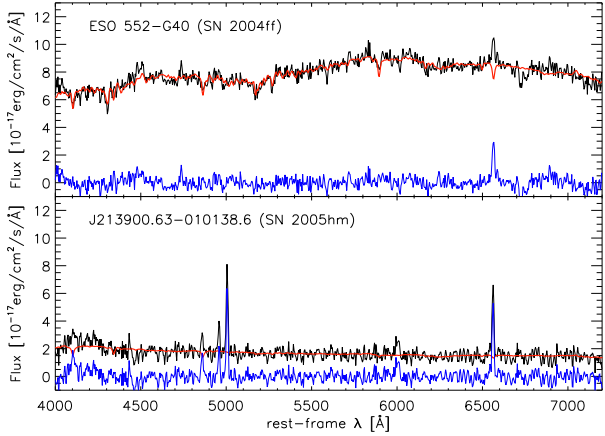}
\caption{
Examples of stellar continuum fit and extraction of emission line spectrum for two representative cases. The black line shows the original spectrum at the location of the SN explosion. This spectrum is fit with a linear combination of SSP models with dust attenuation as an additional free parameter: the resulting best fit of the stellar continuum is shown in red. By subtracting the best-fit continuum from the original spectrum we obtain the pure emission line spectrum (blue line).
}
\label{fig:examspecfit}
\end{center}
\end{figure}

To derive accurate measures of nebular emission lines, which carry information about the gas
metallicity and the age of the youngest stellar populations, care must be taken in accounting for the underlying stellar
continuum and absorption features. This is particularly relevant for the Balmer lines of young stellar populations. To be
able to even recover  weak emission lines, we rely on the procedure implemented in the {\tt platefit} code. This code has been
developed and optimized for the analysis of SDSS spectra, as described in \cite{Tremonti04} and \cite{Brinchmann04}. In
summary, the stellar continuum is recovered by fitting stellar population synthesis models to the galaxy spectrum. The
template model spectra are simple stellar population (SSP) spectra predicted by the \cite{BC03} code at ten different ages between
0.005 and 10~Gyr and three metallicities ($Z=$ 0.2, 1, and 2.5 $Z_{\sun}$, where $Z_{\sun}=0.02$). The templates are first convolved and
rebinned to match the spectral resolution and binning of the observed spectra. The code performs a non-negative least-squares
fit to the emission-line free regions of the spectrum finding the best-fitting linear combination of the template spectra
with dust attenuation as an additional free parameter. The `pure' emission line galaxy spectrum is recovered by subtracting
from the observed spectrum the best-fit continuum and any remaining residuals smoothed on a scale of 200 pixels.
Two representative examples are shown in Fig.~\ref{fig:examspecfit}.

We  measured emission line fluxes and equivalent widths (EW) from the stellar-continuum-subtracted spectra
and  tabulated their values 
in Table~\ref{tab:fluxesNmetal}.    
The fluxes were measured by fitting Gaussians with the IRAF task {\tt splot}.
For [\ion{N}{ii}] $\lambda$6550, H$\alpha$, and
[\ion{N}{ii}] $\lambda$6585, 
we performed  de-blending by fitting simultaneously 3
Gaussians with a single FWHM.  
In Table~\ref{tab:fluxesNmetal}
we have listed the fluxes for
H$\beta$, [\ion{O}{iii}] $\lambda$5008, H$\alpha$, and [\ion{N}{ii}]
$\lambda$6585.
Some galaxies display many more lines,
including [\ion{O}{ii}] $\lambda$3727, [\ion{O}{iii}] $\lambda$4960,
and [\ion{S}{ii}] $\lambda\lambda$6717, 6731;  
however, the tabulated lines 
are those that are more consistently detected and 
those  used for our metallicity determinations. 
The fluxes in Table~\ref{tab:fluxesNmetal} are the ones measured on the emission line spectra (i.e. corrected for stellar absorption).
For reference and to facilitate comparisons with the literature, we also provide the fluxes as measured in the original spectra, both at the host galaxy center and at the SN location (Table~\ref{tab:fluxes2}).

\begin{table*}
\caption[]{Measured line fluxes and H$\alpha$ EW at the SN locations.}
\label{tab:fluxesNmetal}
\centering
\begin{tabular}{lccccc}
\hline\hline
Galaxy &  H$\beta$ &  [\ion{O}{iii}]  $\lambda$5008 &  H$\alpha$ &  [\ion{N}{ii}] $\lambda$6585   &  H$\alpha$ EW    \\	 
\hline
2MASXJ21024677$-$0405233    &      6.08  $\pm$   1.60 &         \mbox{\ldots}        &   29.64  $\pm$   1.70  &   13.32   $\pm$   1.50  &    12.8   $\pm$  0.7  \\
ESO 153$-$G17	      	    &     19.74  $\pm$   4.60 &         \mbox{\ldots}        &   49.83  $\pm$   5.60  &   20.14   $\pm$   2.90  &    20.7   $\pm$  1.5  \\
ESO 552$-$G40	      	    &          \mbox{\ldots}        &         \mbox{\ldots}        &   55.65  $\pm$   7.10  &   13.27   $\pm$   4.00  &  	5.6   $\pm$  0.6  \\
IC 4837A	      	    &          \mbox{\ldots}        &         \mbox{\ldots}        &  145.30  $\pm$   5.60  &   52.02   $\pm$   3.90  &    35.4   $\pm$  1.7  \\
J000109.19$+$010409.5       &     19.21  $\pm$   3.90 &    14.48  $\pm$   4.30 &   45.20  $\pm$   6.30  &    7.40   $\pm$   4.30  &    81.7   $\pm$ 11.9  \\
J001039.34$-$000310.4       &     21.49  $\pm$   4.90 & 	\mbox{\ldots}        &  110.80  $\pm$   6.60  &   43.24   $\pm$   4.70  &    19.7   $\pm$  0.8  \\
J002741.89$+$011356.6       &     28.22  $\pm$   6.90 & 	\mbox{\ldots}        &  121.50  $\pm$   7.70  &   26.22   $\pm$   5.80  &    14.7   $\pm$  0.8  \\
J012314.96$-$001948.8       &     17.28  $\pm$   5.30 &    29.24  $\pm$   5.70 &   57.01  $\pm$   5.90  &   19.93   $\pm$   5.60  &    23.2   $\pm$  2.2  \\
J023239.17$+$003700.1       &    140.80  $\pm$  11.00 &    53.45  $\pm$  13.00 &  634.80  $\pm$  14.00  &  206.70   $\pm$  10.00  &    31.7   $\pm$  0.8  \\
J205121.43$+$002357.8       &    125.40  $\pm$   5.00 &   170.20  $\pm$   6.00 &  317.80  $\pm$   7.40  &   56.93   $\pm$   5.00  &    49.7   $\pm$  1.5  \\
J205519.76$+$003234.4       &    133.20  $\pm$   3.80 &    55.18  $\pm$   4.30 &  512.20  $\pm$  10.80  &  162.70   $\pm$   7.80  &    55.1   $\pm$  2.4  \\
J213900.63$-$010138.6       &     27.83  $\pm$   4.50 &    92.26  $\pm$   4.10 &   87.23  $\pm$   6.10  &    4.07   $\pm$   5.60  &    56.2   $\pm$  4.3  \\
J223529.00$+$002856.1       &    165.20  $\pm$   4.50 &   542.40  $\pm$   4.50 &  431.00  $\pm$   6.40  &   32.59   $\pm$   3.30  &   146.2   $\pm$ 11.7  \\
KUG 2302$+$073	      	    &     49.76  $\pm$   4.60 &    87.38  $\pm$   4.60 &  185.50  $\pm$   6.50  &   29.44   $\pm$   5.00  &    69.3   $\pm$  4.5  \\
MGC$+$03$-$43$-$5\tablefootmark{a}	    &     16.75  $\pm$   2.30 &    11.42  $\pm$   8.80 &   53.71  $\pm$   4.00  &   14.82   $\pm$   2.80  &    11.8   $\pm$  0.7  \\
NGC 1187\tablefootmark{a}    	      	    &     46.39  $\pm$   3.40 &    12.64  $\pm$   2.80 &  176.30  $\pm$   6.50  &   53.03   $\pm$   4.40  &   127.4   $\pm$  6.7  \\
NGC 214\tablefootmark{a}     	      	    &    124.70  $\pm$   9.60 & 	\mbox{\ldots}        &  495.50  $\pm$  14.00  &  163.20   $\pm$   8.10  &    46.1   $\pm$  1.2  \\
NGC 4981    	      	    &     45.37  $\pm$   5.30 &    11.69  $\pm$   4.80 &  259.60  $\pm$   9.30  &  102.70   $\pm$   5.30  &    50.5   $\pm$  2.1  \\
NGC 7364    	      	    &      7.96  $\pm$   3.10 & 	\mbox{\ldots}        &   32.84  $\pm$   6.60  &   18.58   $\pm$   5.90  &    11.0   $\pm$  1.3  \\
NGC 7803\tablefootmark{a}    	      	    &    280.80  $\pm$   6.20 &    51.56  $\pm$   5.50 & 1034.00  $\pm$   8.10  &  464.00   $\pm$   5.50  &    38.7   $\pm$  0.6  \\
\hline\hline
\end{tabular} \\
\tablefoot{
The fluxes are given in units of 10$^{-17}$ erg s$^{-1}$ cm$^{-2}$. The H$\alpha$ EW is in \AA.
\tablefoottext{a}{no sufficient signal was recovered at the SN location and the fluxes correspond to a nearby region (see text for the exact distances).}
}
\end{table*}

\subsection{Metallicity estimates}

To derive metallicities, we  used the empirical O3N2 and N2
calibrations described by \cite{2004MNRAS.348L..59P}.  These methods
present some advantages over other strong-line diagnostics, by being
based on  ratios of neighboring lines and are therefore insensitive to
extinction and to uncertainties in flux calibration.  
Given the limited number of 
[\ion{O}{ii}] $\lambda$3727 detections, combined with the large
uncertainties in the blue part of the spectra, 
no advantage would result from  using  $R_{23}$ or other methods that use this line \citep[see e.g.][]{2002ApJS..142...35K}.
A necessary exception was made in the case of the host of SN~2005hm. 
For this galaxy, the faintest in our sample, [\ion{N}{ii}] $\lambda$6585 was not significantly detected.
Assuming a 2$\sigma$ upper limit for the  [\ion{N}{ii}]  flux yields an upper limit to the (N2) metallicity of $\log\rm{(O/H)} + 12 < 8.44$.
Since [\ion{O}{ii}] $\lambda$3727 is clearly detected for this galaxy and the Balmer decrement is consistent with no host extinction \citep{1989agna.book.....O}, we used $R_{23}$. Taking the average between the calibrations of \cite{M91} (the lower value) and \cite{Z94}, we obtain $\log\rm{(O/H)} + 12 = 8.23$. To account for the different method used and all the related uncertainties, we assigned a conservative uncertainty of 0.20 dex to this value.

The N2 calibrator presents a 1$\sigma$ (2$\sigma$) dispersion of 0.18
(0.41) dex, which is comparable to the dispersion of $R_{23}$-based methods
\citep{2004MNRAS.348L..59P}. An even smaller dispersion of 0.14 (0.25)
dex can be achieved by using the O3N2 method. However, in our case
the determination of O3N2 was possible only for some cases, and for
this reason we focus on the N2 scale in the rest of the paper.  The
estimated metallicities for the local SN environments are tabulated in
Table~\ref{tab:results}.

\begin{table*}
\caption{Metallicities and age estimates at the local SN environment.}
\label{tab:results}
\centering
\begin{tabular}{llccccr}
\hline\hline
Galaxy	& SN & type & O3N2 & N2 & Age (integr.) \tablefootmark{a}  &    Age (min.) \tablefootmark{b}     \\	
	&  &  &  &  & (Gyr)  &   (Myr)     \\	
\hline		           
2MASXJ21024677$-$0405233    &  2007hn  &   Ic	    &	    \mbox{\ldots}	  &    8.70    $\pm$   0.03	&     5.00$_{-1.57}^{+1.81}$	 &     8.6    --   11.1    \\ 
ESO 153$-$G17	      	    &  2004ew  &   Ib	    &	    \mbox{\ldots}	  &    8.68    $\pm$   0.05	&     1.56$_{-0.63}^{+1.12}$	 &     7.3    --    9.7    \\ 
ESO 552$-$G40	      	    &  2004ff  &  Ib/IIb    &	    \mbox{\ldots}	  &    8.55    $\pm$   0.08	&     6.89$_{-1.49}^{+1.60}$	 &    12.3    --   12.7    \\ 
IC 4837A	      	    &  2005aw  &   Ic	    &	    \mbox{\ldots}	  &    8.65    $\pm$   0.02	&     0.92$_{-0.25}^{+0.68}$	 &     6.6    --    6.7    \\ 
J000109.19$+$010409.5       &  2007nc  &   Ib	    &	8.52 $\pm$ 0.10   &    8.45    $\pm$   0.15	&     0.88$_{-0.29}^{+0.67}$	 &     6.3    --    6.8    \\ 
J001039.34$-$000310.4       &  2007sj  &   Ic	    &	    \mbox{\ldots}	  &    8.67    $\pm$   0.03	&     2.81$_{-0.75}^{+1.21}$	 &     7.4    --    9.7    \\ 
J002741.89$+$011356.6       &  2007qx  &   Ic	    &	    \mbox{\ldots}	  &    8.52    $\pm$   0.06	&     2.04$_{-0.59}^{+0.94}$	 &    10.4    --   10.9    \\ 
J012314.96$-$001948.8       &  2006jo  &   Ib	    &	8.51 $\pm$ 0.07   &    8.64    $\pm$   0.07	&     1.89$_{-0.76}^{+1.25}$	 &     7.1    --    9.4    \\ 
J023239.17$+$003700.1       &  2006fo  &   Ib	    &	8.71 $\pm$ 0.04   &    8.62    $\pm$   0.01	&     1.79$_{-0.64}^{+1.68}$	 &     6.7		   \\ 
J205121.43$+$002357.8       &  2007jy  &   Ib	    &	8.45 $\pm$ 0.01   &    8.47    $\pm$   0.02	&     3.24$_{-2.07}^{+1.04}$	 &     7.1    --    7.4    \\ 
J205519.76$+$003234.4       &  2005hl  &   Ib	    &	8.69 $\pm$ 0.01   &    8.62    $\pm$   0.01	&     4.76$_{-1.30}^{+1.72}$	 &     6.3    --    6.4    \\ 
J213900.63$-$010138.6       &  2005hm  &   Ib	    &	    \mbox{\ldots}	  &    8.23    $\pm$   0.20\tablefootmark{c}	&     4.30$_{-2.06}^{+3.04}$	 &     6.9    --    7.2    \\ 
J223529.00$+$002856.1       &  2007qw  &   Ia	    &	8.21 $\pm$ 0.01   &    8.26    $\pm$   0.03	&     2.45$_{-0.73}^{+1.22}$	 &     5.9    --    6.1    \\ 
KUG 2302$+$073	      	    &  2006ir  &   Ic	    &	8.40 $\pm$ 0.03   &    8.44    $\pm$   0.04	&     5.25$_{-2.27}^{+0.04}$	 &     6.6    --    6.9    \\ 
MGC$+$03$-$43$-$5	    &  2005bj  &   IIb      &	8.60 $\pm$ 0.11   &    8.58    $\pm$   0.05	&     0.94$_{-0.23}^{+0.57}$	 &     9.9    --   11.2    \\ 
NGC 1187    	      	    &  2007Y   &   Ib	    &	8.74 $\pm$ 0.03   &    8.60    $\pm$   0.02	&     6.48$_{-2.26}^{+2.64}$	 &     5.9    --    6.0    \\ 
NGC 214     	      	    &  2006ep  &   Ib	    &	    \mbox{\ldots}	  &    8.63    $\pm$   0.01	&     2.04$_{-0.53}^{+0.85}$	 &     6.4    --    6.5    \\ 
NGC 4981    	      	    &  2007C   &   Ib/c     &	8.79 $\pm$ 0.06   &    8.67    $\pm$   0.02	&     2.40$_{-0.68}^{+1.06}$	 &     6.4		   \\ 
NGC 7364    	      	    &  2006lc  &   Ib/c     &	    \mbox{\ldots}	  &    8.76    $\pm$   0.09	&     3.16$_{-1.25}^{+2.14}$	 &     9.9    --   11.7    \\ 
NGC 7803    	      	    &  2007kj  &   Ib/c     &	8.85 $\pm$ 0.02   &    8.70    $\pm$   0.01	&     4.77$_{-1.26}^{+1.55}$	 &     6.6		   \\ 
\hline\hline
\end{tabular} \\
\tablefoot{The tabulated metallicity errors are only the propagated measurement errors. 
To obtain the total uncertainty in the metallicity one needs to add quadratically a systematic error of 0.14 dex for O3N2 and 0.18 dex for N2 \citep{2004MNRAS.348L..59P}.
\tablefoottext{a}{Integrated, luminosity-weighted mean stellar age in the region, obtained by absorption features.}
\tablefoottext{b}{Minimum stellar age, obtained by the H$\alpha$ EW that probes the youngest (most recent) star formation episode.}
\tablefoottext{c}{metallicity computed with the $R_{23}$ method (see text).}
}
\end{table*}

\begin{figure}
\begin{center}
\includegraphics[width=\columnwidth]{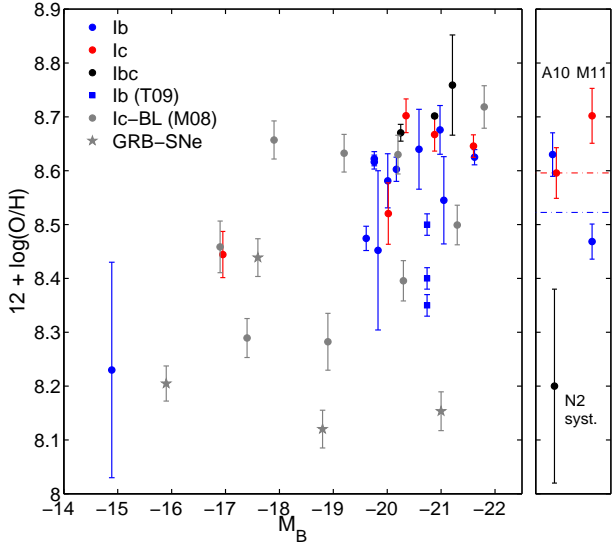}
\caption[]{
Metallicities (N2) at the locations of stripped-envelope CC SNe versus their host galaxy absolute magnitudes. 
SNe~Ib are marked with blue circles \citep[squares for the 3 SNe of][]{CT2770}, SNe~Ic in red and SNe Ib/c (intermediate or uncertain type) in black. Grey stars are for GRB-SNe and gray circles for SNe~Ic-BL \citep{modjaz}.
The displayed error bars only include  the measurement errors, while all data points have an additional associated uncertainty of 0.18 dex, related to the 1$\sigma$ dispersion in the N2 index calibration (shown as the black error-bar in the panel to the right). 
In addition, this panel shows a comparison with the corresponding mean values from \cite{andersonCCmetal} and \cite{modjazIbcMetal}.
Our mean values are marked with dashed-dotted lines.
}
\label{fig:IbIcMetal}
\end{center}
\end{figure}

Plotted in  Fig.~\ref{fig:IbIcMetal} are  the metallicities at the
SN locations versus the host galaxy luminosity.  
In addition to our sample, we have 
also included  the 3 SNe~Ib that occurred in  NGC~2770 for which we conducted 
the measurements in a similar  manner \citep{CT2770}.  
For comparison we  also plotted the GRB associated SNe and the broad-lined
SNe~Ic studied by \cite{modjaz}.  All metallicities were
converted to the N2 scale based on the reported line fluxes
\citep{modjaz,sollerman05,hammer06,Margutti03lw,2007A&A...464..529W}.
The righthand panel shows a comparison with the mean N2 values by
\cite{andersonCCmetal} and \cite{modjazIbcMetal}. 
This is useful to illustrate the corresponding offsets between type~Ib and Ic SNe in the three studies.

The SNe~Ib \citep[$N=14$ including the 3 from][]{CT2770} have a mean
N2 metallicity of 8.52$\pm$0.03 dex.
The quoted error is the \textit{standard error of the mean},
i.e. $\sigma/\sqrt{N}$, where $\sigma$ is the standard deviation.
In this case $\sigma=0.13$.  
The SNe~Ic have a mean metallicity
of 8.60$\pm$0.05 dex and $\sigma=0.11$ on the same scale. This is,
however, based on only $N=5$ events, since
many objects from our initial sample were reclassified.  We note here
that these errorbars (like those  in the mean values quoted by
\citealt{andersonCCmetal} and \citealt{modjazIbcMetal}) are
underestimated because they do not include the uncertainties in the individual
metallicity measurements (largely dominated by the
systematic uncertainties in the metallicity calibrators; see
Sect.~\ref{subsec:discmetal}).

\subsection{Age estimates}
\label{subsec:ages}

We derive two different estimates for the stellar age at the SN location, namely the luminosity-weighted mean
stellar age from the stellar absorption features and the age of the ionizing (youngest) stellar populations from the
H$\alpha$ EW. We measure the stellar absorption features of the spectrum subtracted by the Gaussian fits to the nebular
emission lines. Estimates of the luminosity-weighted mean age are derived by comparing the observed absorption features with
those predicted by a \cite{BC03}-based model library spanning a comprehensive range in random star formation histories and
metallicities, as described in \cite{Gallazzi05}. The constraints are set by a combination of age-sensitive indices (H$\beta$
and H$\gamma_A$+H$\delta_A$) and metal-sensitive indices ($\rm [MgFe]^\prime$ and $\rm [Mg_2Fe]$) to help break the
age--metallicity degeneracy. For each object we derive  the probability density function (PDF) of the luminosity-weighted
age in this way. We take the median of the PDF as the fiducial age estimate and the 16th and 84th percentiles as the 1$\sigma$
uncertainty range. The results are summarized in Table~\ref{tab:results}.

One can immediately see that these ages range between 0.8 -- 7 Gyr, i.e. old compared to the lifetimes of massive stars.
This is expected since the stellar absorption features probe the mean age integrated over the whole star
formation history in the region.
In our case, however, it is more relevant to examine the age of the youngest, i.e. latest,  star formation episode in this region.
The question we
want to address is whether there are any SN locations that, despite the
large uncertainties, indicate a population that is older than what is
allowed by the stellar evolution models for single massive stars.  
Such an age estimator, which is sensitive to the most recent star formation, is the H$\alpha$ EW \citep[see e.g.][]{1999ApJS..123....3L,2001A&A...375..814Z}.
Even so, the relation between H$\alpha$ EW and age strongly depends on the star formation history.
Unfortunately, our question can only be addressed under the assumption that the latest star formation episode was instantaneous, because if star formation  is continuing, massive stars can be born at any time and give rise to CC SNe.
This means that a conclusive answer cannot be given, but we believe that any contribution to the `single versus binary' channel discussion deserves effort. 

\begin{figure}
\begin{center}
\includegraphics[width=\columnwidth]{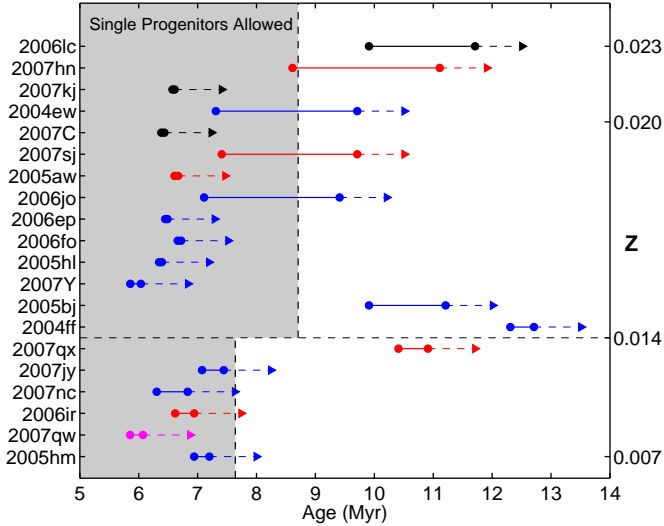}
\caption[]{Ages of the youngest stellar populations at the SN locations, 
estimated through the measured H$\alpha$~EW and Starburst99 \citep{1999ApJS..123....3L}(range denoted by solid line; see text). 
These set a lower limit (dashed line) to the age of the SN birthplace.
The SN environments  were ordered by ascending metallicity, 
and some key values are indicated on the righthand axis, but this is only for illustration purposes as this axis does not have any meaningful scale. 
The color coding is the same as in Fig.~\ref{fig:IbIcMetal}: blue for SNe~Ib, red for SNe~Ic and black for intermediate SNe~Ib/c.
The interloper type~Ia SN~2007qw is in magenta. 
The horizontal dashed line separates the galaxies into two metallicity groups depending on whether a comparison with 
$Z=0.008$ or $Z=0.02$  models  is more appropriate. 
The vertical dashed lines denote the predictions of the Geneva evolutionary codes for the lifetimes of single massive stars of 25 $M_{\sun}$ at $Z=0.02$ and 30 $M_{\sun}$ at $Z=0.008$ \citep{2003A&A...404..975M,MeyMae2005}. 
The allowed ages for single progenitors of SNe~Ib/c are illustrated by the gray-shaded area.
}
\label{fig:ages}
\end{center}
\end{figure}

The ages of the young stellar populations in the vicinity of the SNe have
been estimated by the measured H$\alpha$ EW and by comparison with the predictions of Starburst99 
\citep[][see their Fig.~83]{1999ApJS..123....3L} for instantaneous star formation. 
 Our metallicity estimates were used  to choose the
appropriate table ($Z=0.008$ and $Z=0.02$ are relevant).  
We were 
conservative by providing the widest range in ages
that are compatible with our measurements.  For example, at an
H$\alpha$~EW of $\sim$20~\AA\, at solar metallicity, the models are
degenerate, and many ages between $7-10$ Myr provide a solution.  
We therefore provide a range of possible ages (Table~\ref{tab:results}),
rather than one single interpolated age.  Furthermore, this range is
widened by the uncertainty in our EW measurements and the different
possible IMFs examined by \cite{1999ApJS..123....3L}. 
Indeed, we see that the ages computed this way are much younger than the luminosity-weighted mean stellar ages estimated from the
absorption features and, therefore, represent a \textit{lower limit} to the age of the SN regions.
These lower limits  have been plotted in Fig.~\ref{fig:ages}.
We note that \cite{levesque10grbs} used a similar approach and estimated the ages of the young stellar populations in GRB host galaxies from the
H$\beta$ EW.

\section{Discussion}
\label{sec:disc} 

\subsection{Metallicities}
\label{subsec:discmetal}

Investigating the metallicities of the SN environments is
important because within the single massive progenitor scenario
SNe~Ic are expected to be found in more metal-rich
environments than SNe~Ib.  This is due to the strong dependence of
stellar winds on metallicity \citep{vink}, with stars of the same 
initial mass suffering more severe mass loss and envelope stripping at
higher metallicities.

Our result is intermediate to those of
\cite{andersonCCmetal}, who find equal metallicities for SN~Ib and
Ic environments,  and of \cite{modjazIbcMetal} who find SNe~Ic to lie 
in more metal-rich environments than SNe~Ib. 
\citeauthor{modjazIbcMetal} find a difference of 0.20 dex (in the O3N2 scale) and a 
high probability for rejecting the null hypothesis that   types~Ib and Ic explode in similar metallicity
environments, based on the Kolmogorov-Smirnov (KS) test (p-value = 1\%).
In our sample we cannot detect such a significant difference: a
KS test provides very weak evidence against such a hypothesis (p-value = 17.1\%).

We note that the errors quoted until now (i.e. the standard error of the mean), 
as well as the ones of \citeauthor{andersonCCmetal} and of \citeauthor{modjazIbcMetal},
ignore the relatively large errors in the individual metallicity estimates. 
When the individual error-bars (here $\gtrsim0.18$ dex) exceed
the sample standard deviation $\sigma$ (here $< 0.15$ dex), the
standard error of the mean underestimates the real uncertainty in the
mean value. 
This is the case here because the individual metallicity
errors are dominated by the uncertainty in the N2 (or O3N2) calibration. 
In that case, an error propagation of the individual
errors in the calculation of the mean is more appropriate. 
The same results can be obtained by a simple Monte Carlo (MC) simulation where
the individual metallicities are perturbed around their mean value 
(and within the provided errors) and a metallicity mean is computed in each
realization.  Such a calculation reveals that the true uncertainties
in the mean metallicities are 8.52$\pm$0.05 for SNe~Ib and
8.60$\pm$0.08 for SNe~Ic.
In the rest of the paper all quoted errors are computed with
this method.  
In addition, by performing a KS test for each MC
realization, we find that quoting a single p-value is simplistic but
that there is a distribution of generated p-values with an
associated confidence interval (here 0.007$<p<$0.483 at the 68\%
confidence level). We  therefore find no significant evidence  that types~Ib and 
Ic SNe explode in different environments. 
We believe that taking this into account would also lower the significance of the 0.20 dex difference reported by \cite{modjazIbcMetal}.

It is not obvious why the results from the three  samples differ slightly, especially for the SN~Ib environments.  
Unlike the other studies, we have measured fluxes in spectra where the stellar continuum has been subtracted (Sect.~\ref{sec:results}), but this is not what causes the difference: 
if we use the original spectra (Table~\ref{tab:fluxes2}) we obtain N2=8.54$\pm$0.05 dex for SNe~Ib and 8.61$\pm$0.08 dex for SNe~Ic.
The N2 metallicities are on average a little lower in the stellar-continuum-subtracted spectra (as the corrected H$\alpha$ emission is enhanced), 
but the offset between types Ib and Ic in our sample remains at the same level.
We have compared our results to those that can be obtained on SN spectra that were available to us,  a measurement of metallicity based on lines from the host galaxy was possible (11 cases), and we have found very good agreement. Overall good agreement was also obtained for the \textit{nuclear} metallicities of the five galaxies in our sample that possess an SDSS spectrum. 
At the time this work was accepted for publication, the version of \cite{modjazIbcMetal}  did not provide individual line fluxes or metallicities,
and it was thus not
possible to make direct  comparisons for the five host galaxies that are common to our samples.\footnote{This information is available in the final version of their paper. At this stage, however, it was not possible to include a detailed comparison.}  
\cite{andersonCCmetal} provide metallicities for the individual hosts on the N2 and O3N2 scales, but
there is no overlap between our samples.  
By merging the two samples (using our fluxes in Table~\ref{tab:fluxes2} to avoid differences in methodology), we obtained 8.58$\pm$0.04 dex for SNe~Ib ($N=24$) and 8.60$\pm$0.05 dex for SNe~Ic ($N=19$), i.e. a result intermediate between the two \citep[and qualitatively very similar to][]{andersonCCmetal}.
This, of course, was expected, but we argue below that, at present, a merging of the samples is perhaps not very meaningful.

\begin{table*}
\tiny
\setlength{\tabcolsep}{1pt}
\caption[]{Measured  line fluxes -- not corrected for the effect of stellar absorption.}
\label{tab:fluxes2}
\centering
\begin{tabular}{lcccccccc}
\hline\hline
Galaxy &   \multicolumn{2}{c}{H$\beta$} &  \multicolumn{2}{c}{[\ion{O}{iii}]  $\lambda$5008}  &  \multicolumn{2}{c}{H$\alpha$} &  \multicolumn{2}{c}{[\ion{N}{ii}] $\lambda$6585}        \\	 
\cline{2-3} \cline{4-5} \cline{6-7} \cline{8-9}
   & central & local & central & local & central & local & central & local \\
\hline
2MASXJ21024677$-$0405233   &    	 \mbox{\ldots}	 &	    \mbox{\ldots}	     &  	\mbox{\ldots}       & 	\mbox{\ldots} 	&   148.50  $\pm$   4.30  &    22.68  $\pm$   1.90   &    67.90  $\pm$  3.70  &    11.63  $\pm$   1.50   \\
ESO 153$-$G17	      	   &    	 \mbox{\ldots}	 &	    \mbox{\ldots}	     &  	\mbox{\ldots}       & 	\mbox{\ldots} 	&   898.90  $\pm$  15.00  &    43.37  $\pm$   4.40   &   537.00  $\pm$ 16.00  &    19.13  $\pm$   3.10   \\
ESO 552$-$G40	      	   &    	 \mbox{\ldots}	 &	    \mbox{\ldots}	     &  	\mbox{\ldots}       & 	\mbox{\ldots} 	&	 \mbox{\ldots}	  &    36.75  $\pm$   5.20   &   144.10  $\pm$ 17.00  &    12.27  $\pm$   4.40   \\
IC 4837A     	      	   &    	 \mbox{\ldots}	 &     4.55  $\pm$   1.90    &  	\mbox{\ldots}       &    9.26  $\pm$   5.20	&	 \mbox{\ldots}	  &   125.70  $\pm$   2.20   &   445.70  $\pm$ 12.00  &    47.10  $\pm$   2.10   \\ 
J000109.19$+$010409.5      &     44.22 $\pm$	  7.70   &	    \mbox{\ldots}	     &   39.81 $\pm$   7.10   & 	\mbox{\ldots} 	&   524.90  $\pm$  12.00  &    34.39  $\pm$   5.00   &   190.10  $\pm$  9.30  &     4.90  $\pm$   3.60   \\
J001039.34$-$000310.4      &    	 \mbox{\ldots}	 &	    \mbox{\ldots}	     &  	\mbox{\ldots}       & 	\mbox{\ldots} 	&   450.40  $\pm$  12.00  &    94.81  $\pm$   4.30   &   177.00  $\pm$  8.90  &    35.34  $\pm$   4.20   \\
J002741.89$+$011356.6      &    	 \mbox{\ldots}	 &	    \mbox{\ldots}	     &  	\mbox{\ldots}       & 	\mbox{\ldots} 	&   155.40  $\pm$  10.00  &    85.59  $\pm$   7.60   &    59.48  $\pm$  7.80  &    22.08  $\pm$   5.50   \\
J012314.96$-$001948.8      &     26.34 $\pm$	 11.00   &	    \mbox{\ldots}	     &  131.90 $\pm$  12.00   &   20.21  $\pm$   7.20	&   621.00  $\pm$  16.00  &    27.49  $\pm$   4.60   &   294.30  $\pm$ 12.00  &    15.14  $\pm$   3.60   \\ 
J023239.17$+$003700.1      &     56.45 $\pm$	 13.00   &    41.49  $\pm$  11.00    &   89.82 $\pm$  16.00   &   40.42  $\pm$  12.00	&   960.10  $\pm$  19.00  &   531.30  $\pm$  13.00   &   422.30  $\pm$ 13.00  &   185.10  $\pm$   8.70   \\ 
J205121.43$+$002357.8      &     55.74 $\pm$	  3.90   &    55.74  $\pm$   3.90    &  133.20 $\pm$   5.30   &  133.20  $\pm$   5.30	&   228.80  $\pm$   6.80  &   228.80  $\pm$   6.80   &    42.19  $\pm$  4.00  &    42.19  $\pm$   4.00   \\ 
J205519.76$+$003234.4      &     95.43 $\pm$	  7.10   &    93.31  $\pm$   2.40    &   44.62 $\pm$  12.00   &   45.62  $\pm$   2.80	&  1270.00  $\pm$  15.00  &   477.20  $\pm$   5.40   &   407.80  $\pm$ 11.00  &   162.30  $\pm$   3.90   \\ 
J213900.63$-$010138.6      &     19.05 $\pm$	  3.10   &    19.05  $\pm$   3.10    &   82.52 $\pm$   3.30   &   82.52  $\pm$   3.30	&    78.23  $\pm$   4.30  &    78.23  $\pm$   4.30   &     4.96  $\pm$  6.10  &     4.96  $\pm$   6.10   \\ 
J223529.00$+$002856.1      &    127.10 $\pm$	  3.30   &   127.10  $\pm$   3.30    &  448.80 $\pm$   3.10   &  448.80  $\pm$   3.10	&   390.00  $\pm$   4.80  &   390.00  $\pm$   4.80   &    29.00  $\pm$  3.00  &    29.00  $\pm$   3.00   \\ 
KUG 2302$+$073	      	   &    	 \mbox{\ldots}	 &    35.03  $\pm$   4.20    &  	\mbox{\ldots}       &   79.24  $\pm$   3.80	&	   \mbox{\ldots}	  &   162.80  $\pm$   5.00   &         \mbox{\ldots}        &    26.23  $\pm$   3.30   \\ 
MGC$+$03$-$43$-$5     	   &    	 \mbox{\ldots}	 &	    \mbox{\ldots}	     &   38.63 $\pm$   7.80   & 	\mbox{\ldots} 	&   432.70  $\pm$   9.10  &    41.03  $\pm$   2.70   &   148.00  $\pm$  6.40  &    13.91  $\pm$   2.40   \\
NGC 1187        	   &   3557.00 $\pm$	 34.00   &    41.92  $\pm$   3.40    &  450.20 $\pm$  20.00   &   12.95  $\pm$   2.90	& 25750.00  $\pm$  30.00  &   172.60  $\pm$   6.50   & 12200.00  $\pm$ 24.00  &    51.36  $\pm$   3.60   \\ 
NGC 214          	   &    	 \mbox{\ldots}	 &    75.65  $\pm$   7.50    & 1066.00 $\pm$  26.00   & 	\mbox{\ldots} 	&   507.00  $\pm$  43.00  &   436.00  $\pm$  14.00   &  1528.00  $\pm$ 45.00  &   146.10  $\pm$   8.80   \\
NGC 4981    	      	   &    	 \mbox{\ldots}	 &    24.82  $\pm$   5.50    &  	\mbox{\ldots}       & 	\mbox{\ldots} 	&  2827.00  $\pm$  29.00  &   215.30  $\pm$   7.50   &  1665.00  $\pm$ 23.00  &    88.03  $\pm$   6.30   \\
NGC 7364    	      	   &    	 \mbox{\ldots}	 &	    \mbox{\ldots}	     &  	\mbox{\ldots}       & 	\mbox{\ldots} 	&  1155.00  $\pm$  29.00  &    26.34  $\pm$   3.20   &   718.20  $\pm$ 19.00  &    17.61  $\pm$   3.20   \\
NGC 7803              	   &    	 \mbox{\ldots}	 &   158.90  $\pm$   4.00    &  	\mbox{\ldots}       &   36.65  $\pm$   3.20	&  3450.00  $\pm$  18.00  &   786.80  $\pm$   5.90   &  2250.00  $\pm$ 17.00  &   365.30  $\pm$   3.70   \\
\hline\hline
\end{tabular} \\
\tablefoot{
The fluxes are given in units of 10$^{-17}$ erg s$^{-1}$ cm$^{-2}$ both for the galaxy central regions and for the local SN environments.  
}
\end{table*}

The \cite{andersonCCmetal} sample  has SN classifications  drawn from the Asiago SN catalog, 
and we have demonstrated in Sect.~\ref{subsec:clas} that this can be risky. 
Furthermore, the \cite{andersonCCmetal} sample
is likely to be biased towards higher metallicities because it 
studies SNe that have been discovered in targeted searches of bright nearby galaxies.  
Within our own sample, all CC SNe discovered
by targeted searches ($N=9$) have a mean metallicity of 8.65$\pm$0.06 dex,
while non targeted searches ($N=11$) give 8.56$\pm$0.06 dex (SN~2006lc is
included in both samples). 
In addition, the former seem to cluster
more tightly around their mean (close to solar) value ($\sigma=0.07$), 
while the latter present a much larger dispersion ($\sigma=0.15$). 
This difference is at least as strong as the one found between types~Ib and Ic SNe, and it stems 
 from the galaxy mass-metallicity relation \citep{Tremonti04}.
Also, this is additional evidence that the method of  SN discovery
 is a  constraining factor in the observed environmental properties of SNe.
It is possible that part of the reason  \cite{andersonCCmetal} find relatively higher metallicities is that  
they explored a smaller region of the mass-metallicity parameter space.  
In fact, evidence is mounting that even SN demographics might differ when different types of host
galaxies are probed by different types of searches \cite[e.g.][]{2010ApJ...721..777A}.  
We should therefore be careful before comparing two dissimilar samples.

Concerning  SNe~Ic-BL ($N=9$), on the N2 scale they have 8.51$\pm$0.06 dex
($\sigma=0.16$) and we thus confirm that their environments  are
closer to the SN~Ib distribution than to the normal SN~Ic
distribution \citep{modjazIbcMetal,2010ApJ...721..777A}. Finally GRB-SNe, as pointed out by \cite{modjaz}, are
found in lower metallicity environments (8.23$\pm$0.09 dex;
$\sigma=0.14$ on this scale). The latest nearby event of this kind (GRB~100316D/SN~2010bh) also seems to  follow this picture \citep{2010arXiv1004.2262C,2011MNRAS.411.2792S}.
As a matter of fact, low metallicity has been proposed to be necessary for the formation of a GRB jet \citep{YoonLanger,WoosleyHeger} in an event that would otherwise appear as a regular stripped CC explosion (such as the ones studied in this paper).
Today, however, GRBs have been discovered in high-metallicity environments \citep{Lev020819,2010ApJ...719..378H}, while the fact that we usually detect them at low metallicities can be attributed to a selection bias, such as dust  \citep[e.g.][]{fynbo2009} or star formation rate \citep{2010arXiv1011.4060K, 2011MNRAS.tmp..439M}.

We mentioned earlier that for four galaxies, the measurements were made at distances of 2--7 kpc from the SN (and closer to the galaxy center). Metallicity gradients \citep[e.g.][]{Z94, 1998AJ....116.2805V} make it possible that the metallicities at the actual SN sites are lower. By using the equation derived by 
\cite{BP09}, we estimate that differences ranging between 0.08--0.14 dex should be subtracted by the tabulated values  to obtain the SN location metallicity
for these four galaxies.
Since three of them  are of type Ib (the fourth classified as Ib/c), this will have an implication for the sample mean value, lowering it a little. Indeed, assuming these values are correct, the offset between the types~Ic and Ib metallicity increases to 0.10 dex, becoming slightly more significant. 
We caution, however, that these authors note that metallicity gradients might be flatter than the ones they use  \citep[e.g.][]{2004ApJ...615..228B}
and that this average approach is more suitable for larger statistical samples than for individual galaxies. 
Indeed, for MGC$+$03$-$43$-$5 we did not detect any significant metallicity gradient from the nucleus to the location where the spectrum was extracted
(half way to the SN location).

There is also the question of SN classifications and how this can
affect our samples.  For instance, if SNe~2006lc and 2007kj are
included in the SN~Ib distribution, rather than in the intermediate type~Ib/c sample,
then our estimate for the mean SN~Ib metallicity increases by 0.03 dex.  
We propose that in
the future, as more and better host galaxy and SN data become available,  
such studies should focus on the metallicity dependence of
a continuous SN property, such as the strength of the He lines,
rather than resorting to a bold grouping of either type~Ib or type~Ic SNe. 
Such a methodology is,
however, beyond the scope of the present paper.

Summarizing, based on the present sample and a proper evaluation of the errors, 
we can identify a tentative trend (0.08 dex for our fiducial case) towards, on average, finding 
SNe~Ic  in more metal-rich environments than SNe~Ib, but this is not a significant result.
Given that there is no precise model that quantifies the predicted difference in the mean
environmental metallicities between types~Ib and  Ic SNe, it is difficult to assess the value of this trend. 
Such a quantitative prediction is difficult because the effect of metallicity has to be disentangled from the one of the 
progenitor initial mass, which also has  an observable effect on the types of CC SNe \citep{AndersonJames,K08}, 
and the exact masses of our SNe are not known.
In addition, recent models show that metallicity-driven winds can also play a role in binary evolution at least for some mass ranges  \citep{2010ApJ...725..940Y}. 
Overall, however, the observed trend (if true) does not qualitatively disfavor a single progenitor origin for SNe~Ib/c.

\subsection{Ages}
\label{subsec:agesDisc}

The (minimum) ages of the regions that were derived
in Sect.~\ref{subsec:ages} (with the H$\alpha$ EW method) are all relatively young in an astrophysical
context. They range from $6-13$ Myr, in broad agreement with the
expectation that they should contain massive stars. 

It is possible to make a more quantitative comparison with the
lifetimes of single massive stars that explode as SNe~Ib/c as
predicted by the Geneva evolutionary models
\citep{2003A&A...404..975M,MeyMae2005,Georgy2009}.  
We again used two reference cases, $Z=0.008$ and
$Z=0.02$,  and the horizontal dashed line in Fig.~\ref{fig:ages}
separates our SN environments into two groups, depending on which
metallicity is more relevant for them.  
According to
\cite{Georgy2009}, the lower mass limit above which stars explode as
SNe~Ib/c at $Z=0.008$ is 30 $M_{\sun}$, while this is lowered to 25
$M_{\sun}$ at $Z=0.02$ due to the increased effect of wind mass
loss. The expected lifetimes of stars of these masses 
and metallicities are $\sim$7.6
and 8.7 Myr, respectively\footnote{These are the lifetimes at the end of
core He burning. The remaining lifetimes until core-collapse are however insignificant.}
\citep{2003A&A...404..975M,MeyMae2005}.  
These ages represent 
\textit{upper limits} of the lifetimes of stars that explode as
SNe~Ib/c because more massive stars will explode even  sooner.  
These limits are indicated with the vertical dashed lines in
Fig.~\ref{fig:ages}. We thus expect all SNe~Ib/c that come from the
evolution of single massive stars to be found to the left of these
lines (in the gray shaded area). On the contrary, any SN that exploded outside this area
could have a longer progenitor lifetime and may be due to 
binary evolution (without this possibility being excluded for the SNe in the gray area).
Indeed, WR stars are predicted  to occur over a wider range of ages in stellar populations that include binaries \citep{2009MNRAS.400.1019E}.
We note that despite progress in modeling the evolution of binary stars \citep[e.g.][]{2007PASP..119.1211F,Eldridge2008,2010ApJ...725..940Y},
it is not straightforward to deduce typical lifetimes for systems leading to SNe~Ib/c to make a direct comparison with our data.
The typical timescales depend on a series of parameters and are not as constrained as for single evolutionary models, 
ranging from very young (similar to single stars) to very old.

We observe that the lower limits for the ages of most SN host regions 
are compatible with the single progenitor scenario. 
There are also, however, a few that seem to contradict it.
These are SN~2006lc in NGC~7364, SN~2005bj in MGC$+$03$-$43$-$5, and most
notably, SN~2004ff in ESO 552$-$G40 and SN~2007qx in J002741.89$+$011356.6. 
In addition, the problem is more severe in the case of the He-deficient SNe~Ic that originate 
from even more massive progenitors \citep[39 $M_{\sun}$ at $Z_{\sun}$;][]{Georgy2009}. 
The lifetimes of these stars are even shorter \citep[$<$5 Myr at $Z_{\sun}$;][]{2003A&A...404..975M}, which would also create a problem 
 for the type~Ic SNe~2007sj, 2005aw and, especially, 2007hn. This raises the number of potential discrepancies to 7 out of 19 cases.

As already discussed,  these comparisons have been made against the SN region age \textit{lower limits}.
The areas probed by the slit, however, also contain older stellar populations as is clearly indicated by the higher luminosity-weighted mean stellar ages.
In principle, more SNe can originate in regions that are older than the limits in Fig.~\ref{fig:ages} 
and there could be more explosions attributed to binarity  than in this limiting case.
In addition, these stellar lifetimes were for the Geneva rotating models. Nonrotating models typically
give lifetimes $\sim$15\% shorter for the same stars \citep{2003A&A...404..975M,MeyMae2005}. 
Restricting the comparison to the Geneva models is not a decisive factor since other models give similar (or higher)
ZAMS mass limits leading to SNe Ib/c \citep[34 $M_{\sun}$;][]{2003ApJ...591..288H}
and similar (or shorter) timescales for their evolution \cite[e.g.][]{1993ApJ...411..823W,2003ApJ...592..404L}.

On the other hand, it is true that the quoted discrepancies are small on an absolute scale (between 1--4 Myr) and that 
star formation is present in almost all SN regions we examined 
(with the exception of the 4 SNe where no significant flux was recovered at the exact location to allow us measure an EW). 
That is, we did not find any clear case indicating a very old environment \citep[as e.g.][for SN~2005cz]{2011ApJ...728L..36P}, 
which would unambiguously demand a binary channel explosion (always keeping  in mind these four galaxies).
The most serious concern, however, is that this exercise was made under the assumption of an instantaneous
star formation episode. 
If the SN birthplace  is still forming stars, no such strict limits can be placed through the H$\alpha$ emission.
We have no means of assessing  the validity of such an assumption
(which is indeed challenged by the presence of older stellar generations as indicated by the relatively old mean stellar ages). 
For this reason our conclusions are weak:
we cannot unambiguously claim that the above-mentioned SNe had binary progenitors, but we can propose them as good candidates
and we suggest that this possibility is also investigated by other means (e.g. through the SN
properties).  It is intriguing to point out that two of these explosions
are of type~IIb, i.e. similar to the prototype SN~1993J, the explosion for which the strongest evidence for binarity 
exists \citep[e.g.][]{1993Natur.364..509P,1993Natur.364..507N,2004Natur.427..129M}.

As a parenthesis,  that the lower limit derived for the thermonuclear SN~2007qw environment is as low as 6~Myr 
is no surprise and should not raise any concerns. 
In contrast to CC~SNe, this lower limit does not constrain the age of the SN progenitor in a meaningful way, 
because it is not comparable to the lifetimes of the suspected progenitor systems. 
That SNe~Ia are often associated with young stellar populations ($\sim$50--100~Myr) is well known \cite[e.g.][]{2006MNRAS.370..773M}.

\section{Conclusions}
\label{sec:conc}

We have compiled a sample of 20 well-observed stripped CC SNe. 
Particular attention was paid to classifying  each object,  and
in many cases our spectral typing differs from what
has been reported previously in the literature.  
We  obtained spectra of the host galaxies for these SNe at the 
exact SN location using the NTT ($+$ EFOSC2) and fit these spectra with SSP templates 
in order to estimate and subtract the stellar continuum contribution.

The local metallicities were computed on the N2 scale 
and found to be, after a proper treatment of the
systematic errors,  8.52$\pm$0.05 dex for SNe~Ib and 8.60$\pm$0.08 dex for SNe~Ic.
This may indicate a trend toward increasing metallicity from SNe~Ib to Ic, as expected by single evolutionary models, 
but it is not a statistically significant result.  
A comparison with other studies in the literature was made and differences were discussed.

The ages of the SN environs were estimated by comparing absorption features with population synthesis models (giving a luminosity-weighted mean stellar age) and by the measured H$\alpha$~EW. 
The latter method probes the ionizing youngest stellar populations and was thus used to place lower limits on the ages of the SN birthplaces.
These lower limits were compared with upper limits on the lifetimes of single massive stars as computed by the Geneva evolutionary models.
For a number of SNe (7/19), these limits were found to be mutually incompatible, possibly indicating that they resulted from binary evolution. 
The discrepancies were, however, small (between 1--4 Myr), and all the SN regions we examined included areas of recent ($<$15 Myr) star formation. 
In addition, this conclusion was subject  to the assumption that the SN progenitor was born during an instantaneous star formation episode.

Based on these results, we are not able to conclusively rule out any of the evolutionary paths leading to stripped SN explosions.
The single progenitor channel seems consistent overall  with the observations, while binary evolution might have
been more likely for a few explosions in our sample, due to their associated ages. 
We speculate, in line with many other authors, that stripped SNe probably result from more than one channel. 
Detailed studies of the individual events and their environmnents are needed in order to reveal the nature of each one separately.

\begin{acknowledgements}

We acknowledge helpful comments by Justyn Maund, Christina Th\"one, Erik Zackrisson, and the referee, Phil James.
We thank Linda \"Ostman and Masao Sako for discussions regarding the SDSS-II SN classifications.
We are grateful to Christy Tremonti and Jarle Brinchmann for making their {\tt platefit} code available to us.
GL is supported by a grant from the Carlsberg foundation.
MJM acknowledges the support of the UK Science \& Technology Facilities Council.
The Dark Cosmology Centre is funded by the Danish National Research Foundation.  

\end{acknowledgements}


\bibliographystyle{aa}  
\bibliography{16692}

\end{document}